\documentclass[twocolumn,showpacs,preprintnumbers,amsmath,amssymb,prl,superscriptaddress]{revtex4-1}
\pdfoutput=1
\usepackage{graphicx}
\usepackage{dcolumn}
\usepackage{bbold}
\usepackage{color}

\usepackage[low-sup]{subdepth}
\usepackage{multirow}
\usepackage{amsfonts}
\usepackage[english]{babel}
\usepackage[T1]{fontenc}
\usepackage{times}
\usepackage[colorlinks,bookmarks=true,citecolor=blue,linkcolor=red,urlcolor=blue]{hyperref}
\usepackage[tight, FIGTOPCAP, hang, raggedright, nooneline]{subfigure}

\usepackage{verbatim}
\usepackage{bm}
\usepackage{mathrsfs}

\providecommand{\up}{\uparrow}
\providecommand{\dn}{\downarrow}

\newcommand{\braket}[1]{\left<#1\right>}
\newcommand{\para}[1]{\left(#1\right)}

\newcommand{\COMMENT}[1]{}

\def\beq{\begin{equation}}
\def\eeq{\end{equation}}
\def\barray{\begin{eqnarray}}
\def\earray{\end{eqnarray}}
\def\up{\uparrow}
\def\dn{\downarrow}

\begin{document}
\title{Numerical Observation of Parafermion Zero Modes and their Stability in 2D Topological States}
\author{Mohammad-Sadegh Vaezi}
\affiliation{Department of Physics, Washington University, St. Louis, MO 63160, USA}
\author{Abolhassan Vaezi}
\email{vaezi@stanford.edu}
\affiliation{Department of Physics, Stanford University, Stanford, CA 94305, USA}

\begin{abstract} 
The possibility of realizing non-Abelian excitations (non-Abelions) in two-dimensional (2D) Abelian states of matter has generated a lot of interest recently. A well-known example of  such non-Abelions are parafermion zeros modes (PFZMs) which can be realized at the endpoints of the so called genons in fractional quantum Hall (FQH) states or fractional Chern insulators (FCIs). In this letter, we discuss some known signatures of PFZMs and also introduce some novel ones. In particular, we show that the topological entanglement entropy (TEE) shifts by a quantized value after crossing PFZMs. Utilizing those signatures, we present the first large scale numerical study of PFZMs and their stability against perturbations in both FQH states and FCIs within the density-Matrix-Renormalization-Group (DMRG) framework. Our results can help build a closer connection with future experiments on FQH states with genons. 
\end{abstract}

\maketitle

\begin{figure}[!t]
\centering
\centerline{\includegraphics[width=1.0\linewidth]{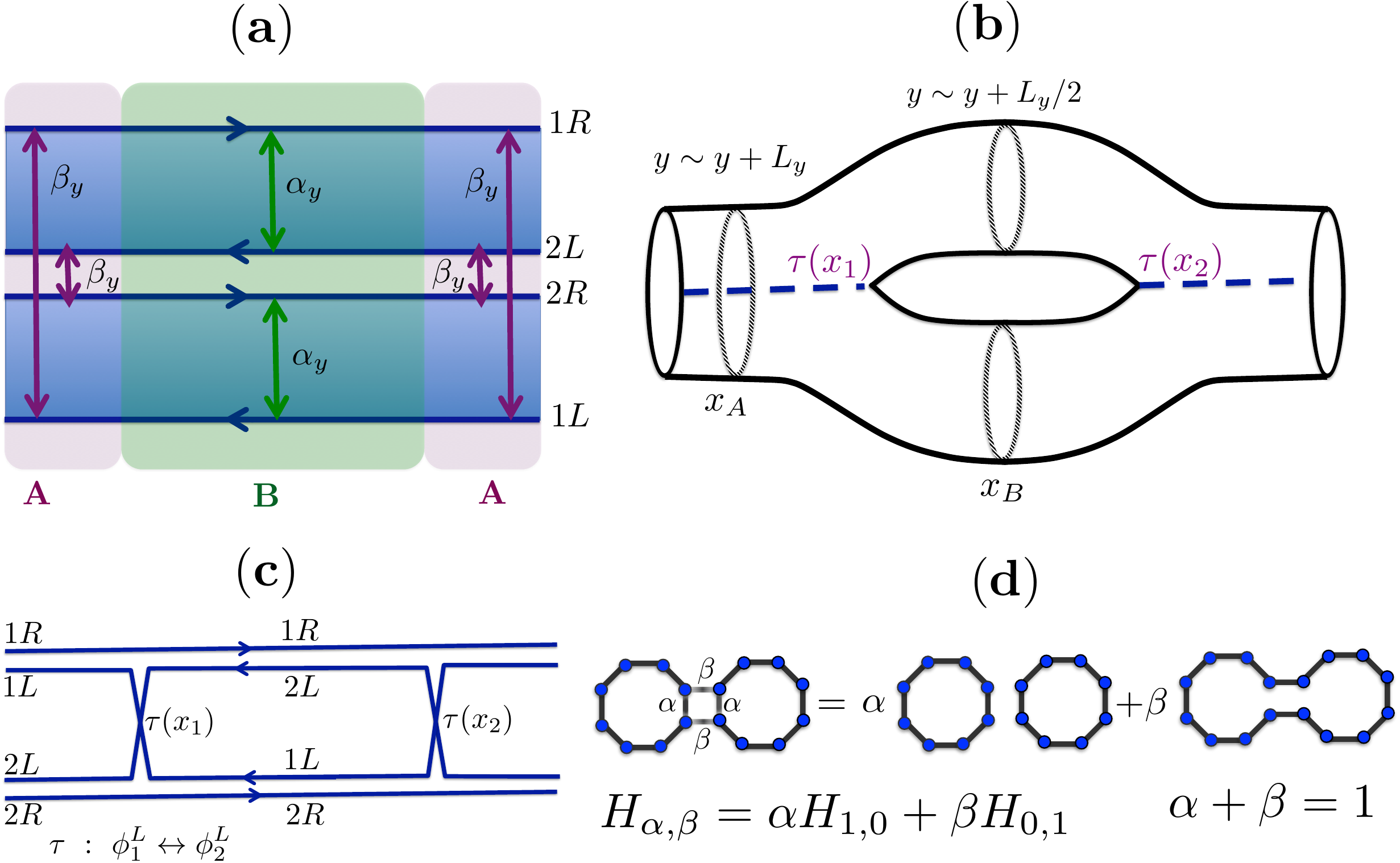}}
\caption{A schematic of genons and the cut and glue approach. (a) Creating two horizontal trenches in topological states results in two chiral and two anti-chiral edge modes. Any two counter-propagating edge modes can be coupled strongly to gap them out which can be viewed as gluing their corresponding edges. In region A, we assume the dominant coupling is between (1R,1L) and (2R,2L) pairs with strength $\alpha_y$ while in region B, the major coupling is turned on between (1R,2L) and (1L,2R) pairs whose strengths are $\beta_y$. We impose the $\alpha_y(x) + \beta_y(x) = 1$ local constraint. (b) The cut and glue procedure will result in Fig. \ref{Fig_00}(b), whose genus is increased by one compared with a case where region B does not exist.  Furthermore, making entanglement cuts at points $x_A$ and $x_B$ will result in one and two homotopically nontrivial cycles, respectively which will impact on the TEE measurement. (c) The juxtaposition of those edges which are interacting more strongly yields Fig. \ref{Fig_00}(c). Doing so, the emergence of a pair of twist operators inserted at $x_1$ and $x_2$ point which exchange the edge modes in one sector only (in this figure, $L$ sector) becomes apparent. We could draw it differently and obtain twist operators exchanging the right modes, while leaving the left modes intact. 
 (d) A cartoon relating the Hamiltonian of ideal systems with a system that is neither a perfect single chain nor  two decoupled chains. This 1d cartoon can be trivially generalized to 2d systems.}
\label{Fig_00}
\end{figure}

\noindent {\em Introduction.--} Non-Abelian (NA) anyons are building blocks of topological quantum computation~\cite{Nayak2008a} and there is an active research on various ways of realizing NA anyons in 2D topological states of matter~\cite{Moore_Read_1991,Greiter_1991,Read1996,qiRMP2011,KaneRMP2010,Fu2008,Lutchyn2010,Kitaev2006a,Yao2007a,Vaezi_2014c,Barkeshli_Jiang_2015,Burrello_genon}. It has been recently shown that NA anyons can be realized in Abelian states through perturbing the parent state by electron pairing, or interlayer electron tunneling/coupling~\cite{wen1991prl,NASS1999,Fradkin1999a,Rezayi_2010a,Qi_Hughes_Zhang_2010,vaezi2013,Vaezi2014a,Mong2014,Vaezi2014b,Oreg2014,Liu_Bilayer_Fib,Geraedts2015a,Peterson2015,Sheng_Zhu_2015,Vaezi_Liu_2016,Repellin_Regnault_2015}. For example, it has been shown that the PFZMs are bound to the endpoints of genons which are domain walls between topologically distinct one dimensional (1D) mass terms (see Fig. \ref{Fig_00}(a))~\cite{clarke2013,lindner2012,Cheng2012a,Barkeshli_Jian2013a,Fendley2012a,Klinovaja2014,Barkeshli2016a}. These PFZMs can be viewed as $Z_N$ generalizations of Majorana zero modes~\cite{You_Wen2012a,bombin2010,beigi2011,clarke2013,lindner2012,Cheng2012a,Barkeshli_Jian2013a,Fendley2012a,vaezi2013,Fradkin_Kadanoff1980a,Ortiz_Cobanera2011a}. Despite the extensive theoretical research on PFZMs~\cite{Barkeshli_2014a,barkeshli2013defect,Teo_genons_rev,Fendley_rev,Teo_genon,Maghrebi2015,Mong_Alicea_PFZM,Santos2017a} and a very recent exact diagonalization study of ideal genons~\cite{Liu_genons}, a careful study of the stability of such NA excitations against perturbations in 2D is still lacking. For examples, it is not clear whether we still obtain PFZMs for a genon where both $\alpha_y$ and $\beta_y$ couplings of Fig. \ref{Fig_00}(a)  are non-zero while their relative strengths yet changes sign. Such questions are relevant for the ongoing efforts towards experimental realization of PFZMs. In this letter, we address this problem in detail and using novel features of PFZMs and large scale DMRG simulations we provide new insights into this problem and present the phase diagram of the system for arbitrary values of $\alpha_y$.

In this letter, we focus on genons in $1/3$ Laughlin state and consider its Landau level (FQH) as well as lattice (FCI) realizations. The genons can emerge in either single layer or bilayer states both with two trenches which lead to two pairs of counter-propagating edge modes localized normal to the boundary of the trenches. The fact that (ideal) genons change the genus of spacetime can be easily understood through what we will refer to as cut and glue approach (or what mathematicians refer to as surgery theory) (see Fig. \ref{Fig_00}(b)). The main idea is that we can create edge modes by making physical cuts (trenches) in the system, and gap them out by strongly coupling a chiral mode with an anti-chiral one. The latter procedure can be imagined as physically gluing those two edges to each other. In Fig. \ref{Fig_00}(a) we have considered periodic boundary condition for the parent state (a single layer $1/m$ Laughlin state) along $y$ direction, thus the modes $1R$, and $1L$ are indeed geometrically neighbors. After digging the trenches and creating the four gapless modes of Fig. \ref{Fig_00}(a), we decide to gap them out again which can be physically achieved by turning on electron tunneling/interaction between counter-propagating edge modes. However, as Fig. \ref{Fig_00}(a) suggests, we have two choices to do so, namely we can pair up edge state (1R) with (1L) (and hence (2R) with (2L) or alternatively (1R) with (2L) (and hence (2R) with (1L)). On the other hand, these two choices are topologically distinct and as a result the domain wall between them carries non-trivial excitations, in this case PFZMs. This result can also be understood within the bosonization approach where the edge theory and their couplings are described by the following sine-Gordon model~\cite{Barkeshli_Jian2013a}:

\begin{eqnarray}
\mathcal{H}^{1D}_{\rm eff} =&&\frac{m}{4\pi}  \int dx \left[\para{\partial_x \varphi_{s}}^2+\para{\partial_x \theta_{s}}^2\right]\cr
-&& g_0\int dx ~\left[\beta_y\para{x} \cos\para{m\varphi_{s}}+\alpha_y\para{x} \cos\para{m\theta_{s}} \right], ~~~~
\label{SG}
\end{eqnarray}
where $\varphi_{s}$ and $\theta_{s}$ are conjugate bosonic variables. In this letter, we assume the $\alpha_y(x) + \beta_y(x) = 1$ local constraint. We now imagine that  $\alpha_y$ is vanishing outside region $B$, and similarly $\beta_y$ vanishes inside $B$. Using the cut and glue approach, we obtain a Riemann surface which is distinct from the parent state which was a cylinder (torus) for open (periodic) boundary condition along $x$. The new surface has an additional hole or genus in the middle, hence the name of genons.  It is worth mentioning that the curvature of the space diverges near $x_1$ and $x_2$ and vanishes away from them, thus only these two points contribute to the Euler character and are responsible for the additional genus. Since the ground-state degeneracy (GSD) of the Abelian states depends on the genus of space manifold~\cite{wenbook}, the GSD changes from $m$ for the parent state (with periodic boundary condition (PBC) along $x$ direction) to $m^2$ for the perturbed state. The synthesized genon is solely responsible for the additional $m$-fold degeneracy, and since it has two endpoints ($x_1$ and $x_2$), we expect a zero-mode operator with quantum dimension of $d_m = \sqrt{m}$ (i.e., the advertised $Z_m$ PFZMs) at each endpoint. 
So far, we only considered the ideal case where $\alpha_y(x)\beta_y(x) = 0$ everywhere. Now, we may ask what if we have a situation where $\alpha_y(x)\beta_y(x)$ is finite everywhere? Do we still expect PFZMs at the domain walls? How protected are they, and similar questions. We will study these questions carefully in the remainder of this letter.

Before proceeding, we would like to point out a few remarks regarding the confinement of the PFZMs and genons. If an excitation can be achieved without changing the Hamiltonian, it is called a dynamical or a deconfined excitation such as fractional charge excitations of Laughlin states. However, if an excitation can only be achieved through (usually locally) changing the Hamiltonian itself first (or imposing a certain boundary condition), it is called a confined excitation, e.g., the Majorana zero modes bound to vortices in topological superconductors which require creating vortices first~\cite{Read2000,Ivanov2001}. Genons can be viewed as changing the Hamiltonian by inserting a pair of twist operators which change the boundary condition along $y$ direction on one side and cause singularity in the local curvature of the space manifold around them. 
Therefore, genons and PFZMs bound to them are confined anyon excitations. This can also be understood by noting that twist operators act on the bosonic edge modes through either exchanging the two right-moving or left-moving ones (see Fig. \ref{Fig_00}(c)). In the first situation, their topological spins (related to their conformal dimensions) are positive while in the second case they would be negative. Both choices are valid and the resulting chiral and anti-chiral twist operators must be identified accordingly. However, it is known that identifying two anyons with opposite topological spins leads to anyon confinement~\cite{bais2009}.

\begin{figure}[!t]
\centering
\centerline{\includegraphics[width=1.0\linewidth]{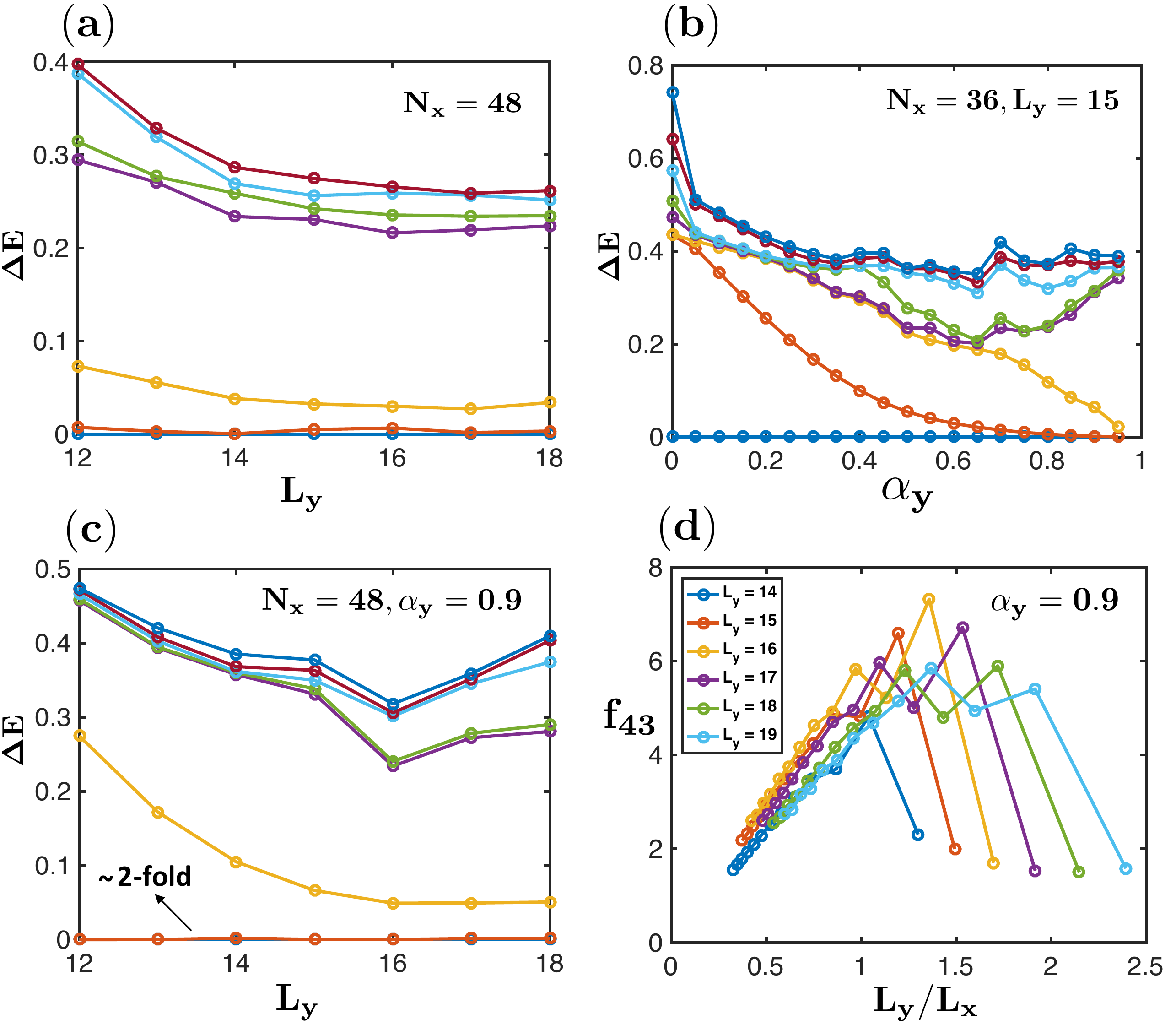}}
\caption{The lowest energy eigenvalues (relative to the ground-state) in one of the three topological sectors for FQH models. (a) Energy spectrum of a finite horizontal genon between $\alpha_y = 0$, and $\alpha_y = 1$ versus $L_y$. (b-c) Energy spectrum of infinite horizontal genons versus $\alpha_y$ (b), and $L_y$ (c). (d) $f_{34} = (E_4-E_1)/(E_3-E_1)$ against aspect ration $(L_y/L_x)$. These plots imply that there is a topological phase transition near $\alpha_y = 1/2$ in the thermodynamic limit, where the GSD changes from 3 to 9 via increasing $\alpha_y$.}
\label{Fig_01}
\end{figure}

In the following, we consider two types of systems to study the stability of genons. (a) Haldane model based FCI studied in Refs. \onlinecite{neupert2011,Grushin2015a}. In this case, we have genons and trenches with sharp edges in real space. (b) FQH system, where we consider genons extended horizontally and in some cases vertically. For these cases, in order to enjoy momentum conservation and avoid Landua level mixing, we assumed genons with smooth boundaries which are still rich enough for our purposes. For FQH based models of genons with sharp boundaries in real-space see the Supplemental Material (SM). In the following, by finite genons we mean genons that are extended between $x_1$ and $x_2$ points, and the values of $\alpha_y$ and $\beta_y$ are exchanged upon crossing these points. Infinite genons means $x_1$ and $x_2$ are pushed to the boundaries of the system. As Fig. \ref{Fig_00}(d) suggests, the Hamiltonian of a region with arbitrary $\alpha_y$ and $\beta_y$ couplings is a linear combination of the two Hamiltonians associated with $\alpha_y = 0$, and $\beta_y = 0$, respectively. In this paper, $N_x$ denotes the number of orbitals in the lowest Landau level. We have considered systems up to 144 electrons and DMRG calculations~\cite{White1992a,McCulloch2008a,Zaletel2012,Zaletel_IDMRG_a} with bond dimensions up to 4000 states. We have first run infinite DMRG algorithm introduced in Ref. \onlinecite{McCulloch2008a} and finally made several sweeps with finite DMRG algorithm (see the SM for more details about the models). We use the following various diagnostics to study the stability of genons and their PFZM bound states against inter-edge couplings:

\noindent {\bf (1) Ground-state degeneracy (GSD):} As discussed above, the quantum dimension of the $Z_m$ PFZMs is $\sqrt{m}$, so the GSD in the presence of $2n$ PFZMs would multiply by $m^n$. Fig. \ref{Fig_01}(a) shows the energy spectrum in one topological sector of a $1/3$ Laughlin state with a single finite genon (extended between $x= \pm L_x/4$) and PBC for several parameters. We have employed the method introduced in Ref. \onlinecite{Vaezi_Entanglement_Distance} to obtain the GSD with DMRG. The given GSD must be multiplied by three to obtain the overall GSD as there are three topological sectors in the parent state. The lowest three energy eigenvalues in the chosen topological sector are quite close compared with higher energy states. The finite splitting is due to the finite size effect, which becomes more evident by plotting energy spectrum as a function of $L_y$ and $N_x$. Another interesting fact is that the ground-state energy is nearly independent of the system size implying the localized nature of PFZMs (see the SM for more details). In Fig. \ref{Fig_01}(b-d) we consider infinite horizontal genons (extended between $x = \pm L_x/2$) for various parameters. As it can be seen, the GSD is $1 (\times 3)$ for $\alpha_y=0$ and becomes $3 (\times 3)$ for $\alpha_y = 1$. Therefore, there should be a quantum critical point in between. Using various tools, we show that the phase transition happens near $\alpha_y =1/2$. In the thermodynamic limit, the GSD is expected to be $3  (\times 3)$-fold for $\alpha_y > 1/2$, while there would be finite splitting due to the finite size effect. To gain a better insight on the finite size effect, we plot the $f_{43} \equiv \frac{E_4-E_1}{E_3-E_1}$ quantity against the aspect ratio ($L_y/L_x$) for various system sizes. As Fig. \ref{Fig_01}(d) suggests, $f_{34}$ takes it maximum around $r \sim 1$. Finally, we obtain similar results for vertically extended genons (see the SM). 

\begin{figure}[!t]
\centering
\centerline{\includegraphics[width=1.0\linewidth]{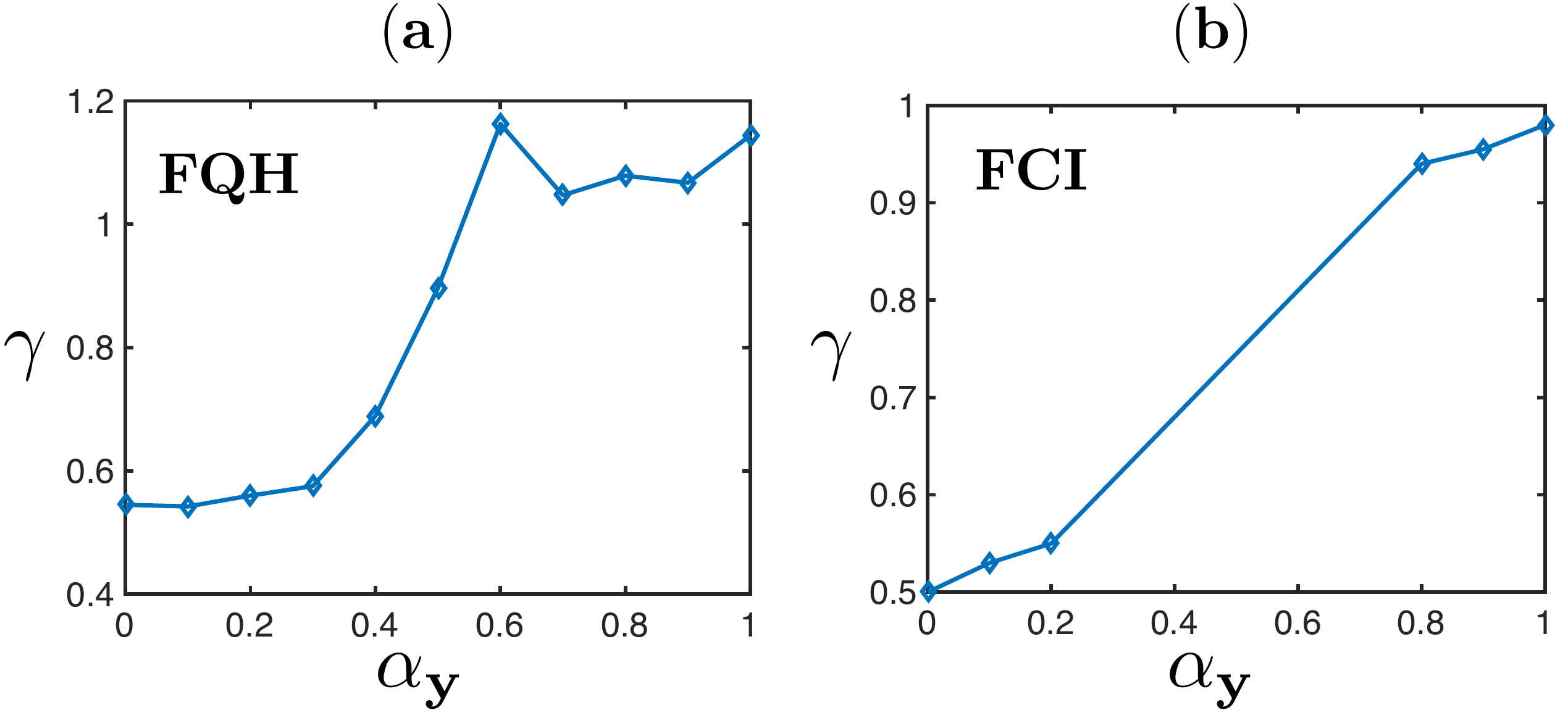}}
\caption{TEE ($\gamma$) measured as a function of $\alpha_y$ for FQH (a) and FCI (b) models on cylinder. We predict that $\gamma$ is $\gamma_0 = \log(\sqrt{3}) \simeq 0.549$ for OBC when $\alpha_y < 1/2$ and in the thermodynamic limit, while it is $2\gamma_0$ for $\alpha_y > 1/2$. The finite size effect however enlarges the critical point to a finite window around $\alpha_y =1/2$.}
\label{Fig_02}
\end{figure}

\noindent {\bf (2) TEE shift:} Let us consider a horizontal genon. As illustrated in Fig. \ref{Fig_00}(b), the entanglement cuts before and after the endpoints of a genon are topologically distinct, namely for open boundary condition (OBC) the cut at $x_A$ yields a single fictitious edge state of length $L_y$ and for $x_B$ we obtain two fictitious edge states each defined on a chain of length $L_y/2$. Accordingly, the TEE is ~\cite{Kitaev2004a,Levin2004a,Zozulya2007,Dong_Fradkin2008a} $\gamma_0 = \log(\sqrt{m})$ on the left side of the $x_1$, and $2\gamma_0 = \log(m)$ for the middle points. Thus, the TEE shifts by $\gamma_0$ when crossing PFZMs bound to the twist operators. In other words, the TEE is $2\gamma_0$ for $\alpha_y > 1/2$ and $\gamma_0 $ for $\alpha_y < 1/2$ regions for OBC. Another way to reach this result is by noting that the TEE of an edge state with total anyon charge $a$ is $\log(\mathcal{D}/d_a) $, where $\mathcal{D}$ is the total quantum dimension and $d_a$ the quantum dimension of $a$. Since for these twist operators, $d_{\tau} = \sqrt{m}$, and a region with $\alpha_y < 1/2$ encloses one $\tau$ operator while the region with $\alpha_y > 1/2$ does not, the TEE shift is equal to $\log(d_{\tau}) = \gamma_0$. 
In Fig. \ref{Fig_02} we have studied this entanglement shift for both FQH and FCI states. We can clearly see the TEE shift between $\alpha_y < 0.4$ and $\alpha_y > 0.6$, while due to finite size effect the TEE shift for the intervening parameter regime is harder to achieve.

\begin{figure}[!t]
\centering
\centerline{\includegraphics[width=1.0\linewidth]{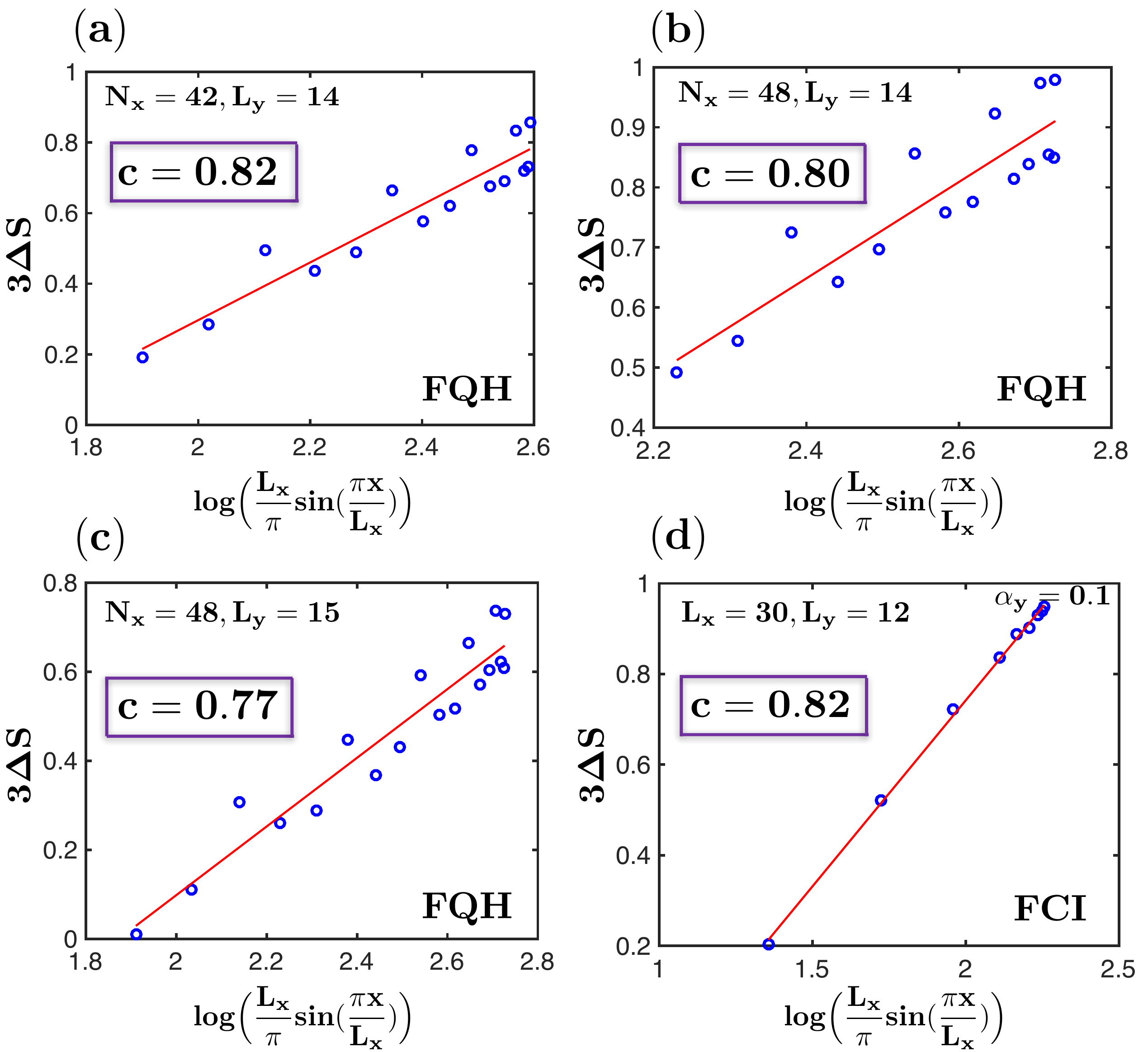}}
\caption{The von Neumann entanglement entropy versus $x$ at the critical point for FQH  (a-c) and FCI (d) models for various circumferences. The measured central charge is close to its theoretical value, $4/5$ as critical genons for 1/3 filling are described by $Z_3$ parafermion CFT. }
\label{Fig_03}
\end{figure}

\noindent {\bf (3) Central charge measurement:} The $Z_m$ sine-Gordon model of Eq. \eqref{SG} is self-dual for $\alpha_y = 1/2$, and hence is expected to be critical. For the $m=3$, it is known that it flows to $c=4/5$ parafermion conformal field theory (CFT) in the IR limit~\cite{lecheminant2002}. Therefore, the whole system can be imagined as a gapped 2D system in addition with a 1D critical line~\cite{Zaletel2013edge}. Thus, the entanglement entropy for an entanglement cut at $x$ must behave as $S_{\alpha_y=1/2}  = S_{\alpha_y=0} + S_{CFT} $, where $S_{\alpha_y = 0} \to aL_y -\gamma$ for $x\gg l_B$, and $S_{CFT} = n_0 \frac{c}{6} \log(\frac{L_x}{\pi} \sin(\frac{\pi}{L_x} x))$ where $n_0 = 1$ (2) for OBC (PBC)~\cite{Calabrese2004a}. Therefore, we can read the central charge of the critical genons by measuring $S_{\alpha_y=1/2}-S_{\alpha_y=0}$. In Fig. \ref{Fig_03} and in the SM we show the result of the entanglement calculation for various parameters for FQH and FCI. We see a remarkable agreement between theory and numerics.

\begin{figure}[!t]
\centering
\centerline{\includegraphics[width=0.5\linewidth]{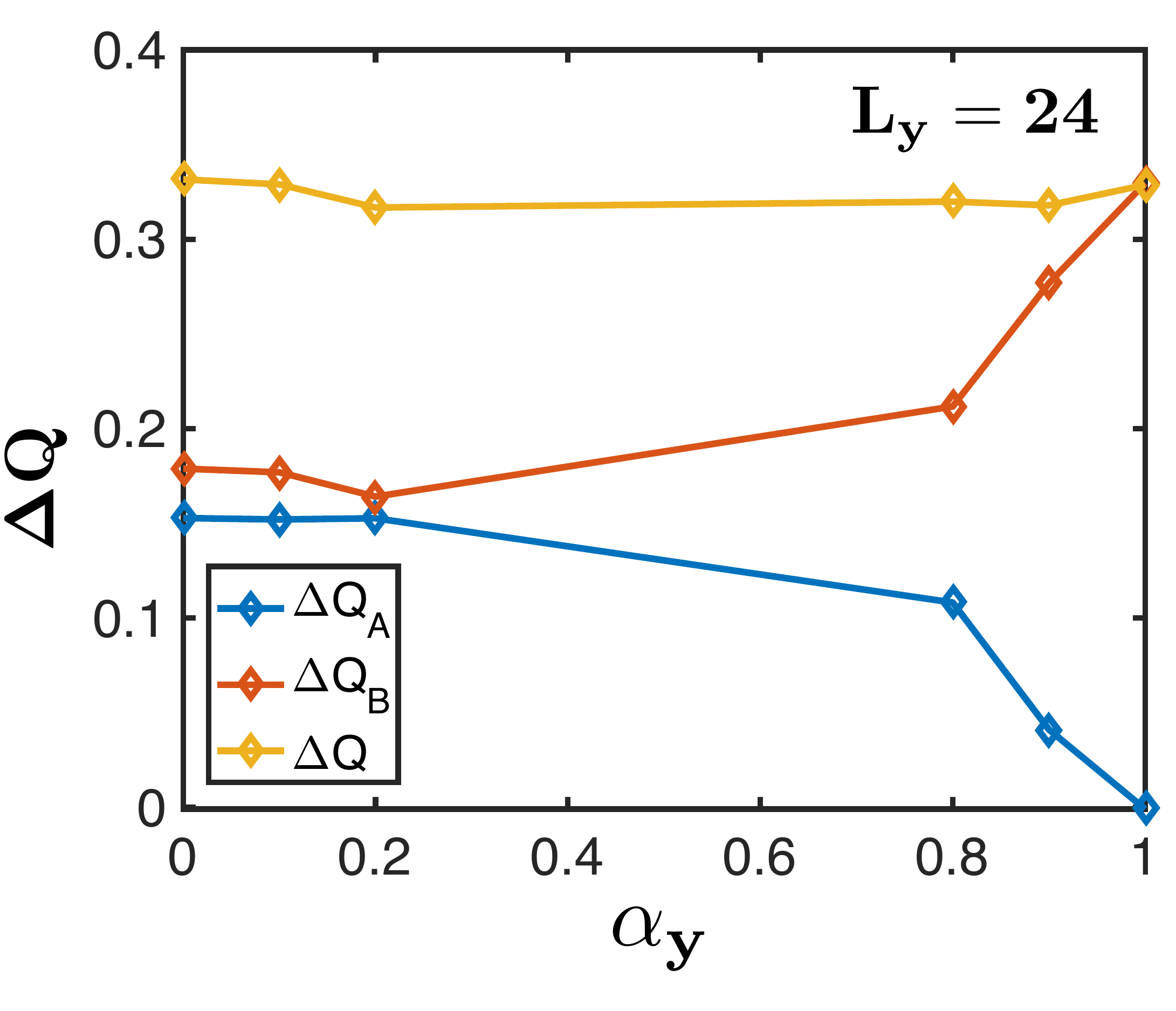}}
\caption{Charge pumping due to one quantum of flux insertion versus $\alpha_y$. $\Delta Q_A$ ($\Delta Q_B$) denotes the charged pumped through the lower (upper) cylinder (in our model odd and even orbitals, respectively), and $\Delta Q$ the sum of the two. We see that $\Delta Q_A \simeq \Delta Q_B$ for $\alpha_y < 1/2$, while for $\alpha_y > 1/2$, one cylinder is mostly responsible for the charge transport.}
\label{Fig_04}
\end{figure}

\noindent {\bf (4) Charge pumping and flux insertion:} Inserting a flux $\phi$ in a topological phase on the cylinder
will pump $\sigma_{xy}\phi$ electric charge from one boundary to the other one~\cite{wenbook,Grushin2015a,He2014a,Vaezi_Hafezi_2016}. For the $1/m$ Laughlin state, we can label the $m$ distinct topological sectors on the cylinder by their $q_n=ne/m$ (mod 1) fractional charges on the left boundary, where $n=0,\cdots,m-1$. As a result, inserting a quantum of flux adiabatically takes us from ground-state $n$ to ground-state $n+1$ (mod $m$). 
Now, consider a horizontal genon with $\alpha_y = 0$. We like to insert flux $\phi$ in the system which can be done unambiguously as the system can be viewed as a single cylinder. However, there is an ambiguity in the determination of the flux threading the upper ($\phi_{\up}$) or lower ($\phi_{\dn}$) branches of a genon with $\alpha_y = 1$, since we are now dealing with two decoupled cylinders  (i.e. $y \leq L_y/2$ and $y > L_y/2$ regions). The only constraint is that  $\phi_{\up} + \phi_{\dn} = \phi$, which fixes one of them only. Thus, $\phi_{\up}$ can take any of the $m$ possible quantized values, leading to an additional $m$-fold degeneracy of the whole system. In terms of the charge polarization (transport), measuring the amount of charge pumped through the upper branch must exhibit quantization in units of $n_{\up}e/m$ for $\alpha_y = 1$. On the other hand, for $\alpha_y = 0$,  symmetry dictates the two (fictitious) cylinders to contribute equally to the charge pumping in the thermodynamic limit. Thus,  the charge pumped through each cylinder is $n e/2m$ rather than $ne/m$. Generalizing this observation to arbitrary $\alpha_y$, we predict that the quantum of pumped charge is $ne/2m$ (~$ne/m$~)  for regions with $\alpha_y < 1/2$ ($\alpha_y> 1/2$). Fig. \ref{Fig_04} summarizes our results for the charge pumping measurement.

\begin{figure}[!t]
\centering
\centerline{\includegraphics[width=1.0\linewidth]{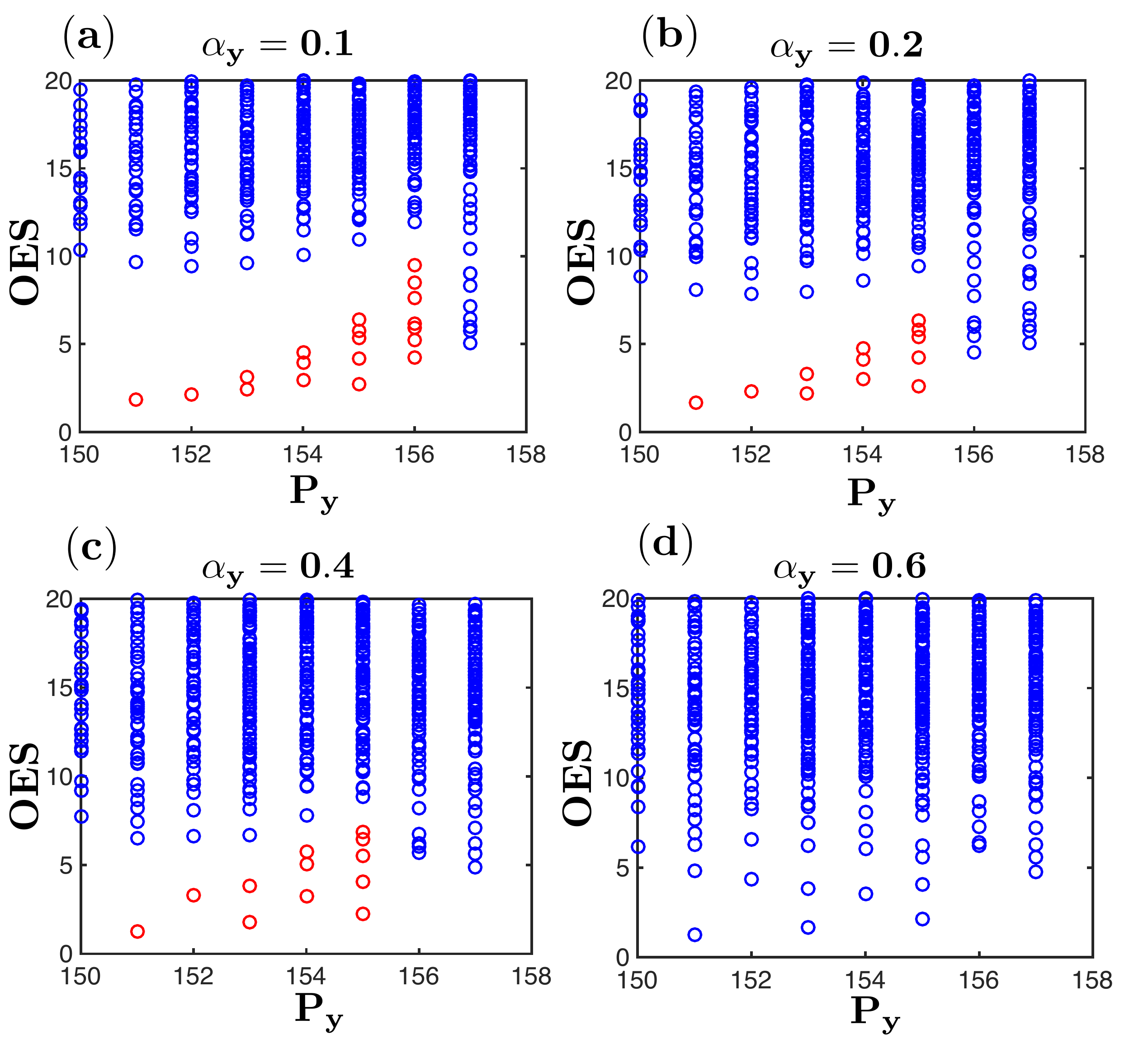}}
\caption{OES versus $\alpha_y$. We predict the low lying edge-like excitations to follow $1,1,2,3,5,7,\cdots$ counting for $\alpha_y < 1/2$. This figure shows a remarkable agreement between theory and numerics. Due to the finite size effect, $\alpha_y = 0.6$ is nearly critical, and the gap between the edge-like and bulk spectra vanishes.}
\label{Fig_05}
\end{figure}

\noindent {\bf (5) Orbital Entanglement Spectrum (OES):} The physical edge states of topological states have a characteristic degeneracy for a given momentum on the cylinder geometry. For example, the edge state of a single layer FQH follows $1,1,2,3,5,7,\cdots$ counting for $\Delta Q = 0,1,2,3,4,5,\dots$ momenta~\cite{wenbook}. A similar trend is expected for entanglement cut~\cite{Li2008,Thomale2010}. Now, consider a region with $\alpha_y < 1/2$, which is smoothly connected to $\alpha_y = 0$. Such a region can be viewed as a single cylinder, and the counting must follow $1,1,2,3,5,\cdots$, while a region with $\alpha_y > 1/2$ must exhibit a different counting consistent with two decoupled cylinders. In Fig. \ref{Fig_05}, we show the robustness and power of the OES versus $\alpha_y$ for detecting the quantum phase transition.

\noindent {\bf (6) Momentum Polarization (MP):} It is shown by Tu et al ~\cite{Qi_momentum_polarization} that translating a region near one boundary of a topological state along $y$ direction can detect the central charge of the edge state as well as the topological spin of the anyonic excitations on that boundary. This is due to the fact that the boundaries are gapless, and the translation operator is (partially) related to the computation of the CFT partition function in a given topological sector (primary field) at zero temperature which is proportional to $\chi_a\para{L_y} = \exp\para{2\pi i\para{h_a-c/24}/L_y}$. From this simple fact, we can easily compute the topological spin of the twist operators living at the endpoints of a genon. To this end, we need to compare $\chi_{0}\para{L_y}$  corresponding to an $\alpha_y =0$ region with trivial charge, with $\chi_{0}\para{L_y/2}\chi_{0}\para{L_y/2}$, which correspond to a region with $\alpha_y =1$ (we could indeed assign $a_1$ and $a_2$ anyon charges as long as $a_1+a_2 = 0$). The twist operator compensates the difference between the two regions, and thus $\chi_{\tau} = \exp\para{2\pi i h_{\tau}/\para{L_y/2}}$, where $h_{\tau} = c/16$. 
This is consistent with predictions relating PFZMs to the twist operators of an orbifold $U(1)_m$ CFT whose conformal and quantum dimensions are $1/16$ and $\sqrt{m}$, respectively \cite{bais2009,vaezi2013}. Testing this prediction for FQH states numerically requires regions with sharp boundaries in real space, while our computations for FQH states are done in orbital space whose boundaries are smooth in real space.

\noindent {\em Conclusion.--} We presented the first large scale numerical results for the topological geometric defects in FQH states and studied the stability of the emergent NA anyons using state-of-the-art DMRG algorithm. We showed that creating genons changes the GSD not only for the ideal case where $\alpha_y\beta_y =0$, but also when both couplings are nonzero. We listed a number of other novel signatures that can be employed to study the stability of genons and the PFZMs bound to them and characterize the topological phase transition. We carried our computations on both lattice realization of topological states (FCI) and FQH systems. Our model Hamiltonian for FQH is simple and captures the essential physics of the genons, and can be generalized to study genons in more exotic phases. There are still several interesting questions that need to be investigated carefully. For example, the time evolution of the system can shed light on the braid statistics of PFZMs. Also, the critical genons have a different CFT description from the bare edges. The detailed knowledge of the transmutation of the anyon content and primary fields of the two sides is still lacking and future numerical studies can be helpful in this regard.

\noindent{\em Acknowledgements.--} We acknowledge helpful discussions with Xiao-Liang Qi, Srinivas Raghu, Zohar Nussinov, Hongchen Jiang, Yin-Chen He, Prashant Kumar, Chao-Ming Jian, Seyed Mahmood Hoseini and Sai Iyer. We thank XL Qi and S Raghu for kindly granting us access to their computing workstations for our numerical simulations. AV was funded by the Gordon and Betty Moore Foundation's EPiQS Initiative through Grant GBMF4302. MSV was partially supported by the National Science Foundation under NSF Grant No.DMR-1411229.

%

\begin{widetext}
\newpage
\appendix

\section*{ SUPPLEMENTAL MATERIAL:\\
 Numerical Observation of Parafermion Zero Modes and their Stability in 2D Topological States}

In this supplemental material, we provide more details on the model Hamiltonians used in the main text and present additional numerical results for the interested reader.

\subsection*{A. Our model Hamiltonian of genons for FQH state}
In this section we present our model Hamiltonian of genons for the Landau level realization of Laughlin state, and also comment on a related model which sharp boundary trenches.  Let us first consider a chain of Fig. \ref{Appendix_Fig_01}. The Hamiltonian of this chain in real space can be written as

\begin{eqnarray}
H_{\alpha,\beta} = \alpha H_{1,0} + \beta H_{0,1},
\end{eqnarray}
where $H_{0,1}$ is the Hamiltonian of a single chain of length $L_y$, while $H_{1,0} = H_A \oplus H_B$ is the Hamiltonian of two decoupled $A$ and $B$ chains each of length $L_y/2$. For later convenience, we assume that chain $A$ is subject to the periodic boundary condition, while chain $B$ is subject to the anti-periodic boundary condition.  Therefore, the momentum eigenvalues of chain $A$ are quantized as $p^A_k = \frac{2\pi k}{\para{L_y/2}}$, while for $B$ we have $p^B_k = \frac{2\pi \para{k+1/2}}{\para{L_y/2}}$. Additionally, the momentum eigenvalues of the larger chain are quantized as $p_k = \frac{2\pi k}{\para{L_y}}$. The form of $H_{\alpha,\beta}$ in momentum space is rather involved, since the momentum states of the smaller chains are nontrivially related to those of the larger chain. In order to see their relations, we can use the following expressions for the electron annihilation operators:

\begin{eqnarray}
&&c_{k} = \sum_{y=1}^{L_y}F(y,k)c_{y} = \sum_{y = 1}^{L_y/2}F(y,k) c^A_{y} + \sum_{y = L_y/2+1}^{L_y}F(y,k) c^B_{y} \\
&& F(y,k) = \exp\para{ip_k y}/\sqrt{L_y},\quad p_k = 2\pi k/\para{L_y}.
\end{eqnarray}

We can consider the even and odd momenta separately after which we obtain:
\begin{eqnarray}\label{genon2}
&& c_{2k} = \frac{c^A_{k} + c^B_{k+1/2}}{\sqrt{2}}\\
&& c_{2k+1} = \frac{c^A_{k+1/2} - c^B_{k}}{\sqrt{2}},
\end{eqnarray}
relation. However, as mentioned earlier, the momentum eigenvalues of chain $A$ are quantized as $4k\pi/L_y$, thus the state with $4(k+1/2)\pi/L_y$ momentum which does not exist among them must be a linear combination of all valid momentum states. Similarly, for chain $B$, momenta are naturally quantized as $4(k+1/2)\pi/L_y$, and therefore $4k\pi/L_y$ must be expandable in terms of valid momentum states. To find these linear relation, we can first to relate the real space electron annihilation operators of chains $A$ and $B$ to their momentum counterparts as follows:

\begin{eqnarray}
&&c^A_{y} = \sum_{p_A = 4\pi k/L_y}F^*_A(y,k)c^A_{k}\\
&&c^B_{y} = \sum_{p_B = 4\pi \para{k+1/2}/L_y}F^*_B(y,k)c^B_{k+1/2}\\
&& F_A(y,k) = \exp\para{ip_A y}/\sqrt{L_y/2},\quad p_A = 4\pi k/L_y\\
&& F_B(y,k) = \exp\para{ip_B y}/\sqrt{L_y/2},\quad p_B = 4\pi \para{k+1/2}/L_y.
\end{eqnarray}

Plugging the above relations into Eqs. \eqref{genon2}, we obtain the following relations:

\begin{figure}[!t]
\centering
\centerline{\includegraphics[width=.5\linewidth]{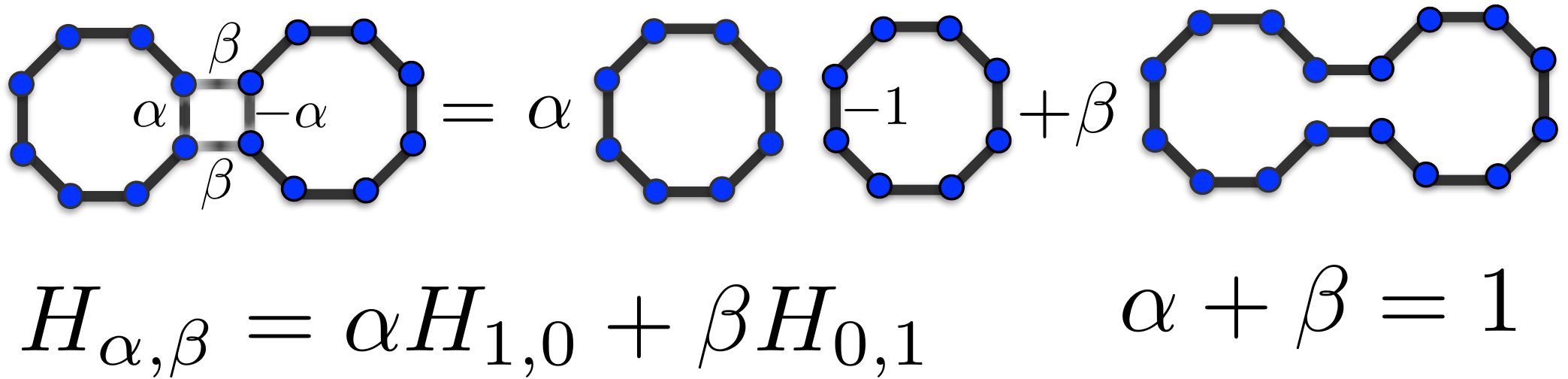}}
\caption{A cartoon relating the Hamiltonian of a chain that is neither a single large periodic chain nor two decoupled smaller periodic and anti-periodic chains to the ideal cases. For later convenience, we have imposed the anti-periodic boundary condition on one of the small chains. }
\label{Appendix_Fig_01}
\end{figure}

\begin{eqnarray}
&& c^A_{k+1/2} = \sum_{q} U_{k}^q c^A_q\cr
&&U_k^q =  \frac{2}{1-exp\para{i\frac{4\pi }{L_y}\para{k+1/2-q}}}
\end{eqnarray}
and a similar relation for $c^B_k$.

Now, let us consider the Landau level problem. Similar to the above relations, we need to first expand the electron annihilation operator with momentum $2\pi (2k)/L_y$ and $2\pi (2k+1)/L_y$ and on the $n$-th Landau level to their real space counterparts after which we arrive at:

\begin{eqnarray}
&&c_{k,n} = \sum_{x,y}F^*(n,p,x,y)c_{x,y},\quad p = 2\pi k/L_y \\
&&F(n,p,x,y) = H_n(x-pl_B^2) \exp(ipy)/\sqrt{L_y},
\end{eqnarray}
where $H_n(x)$ is the $n$-th Hermite's functions (we use the convention where Hermite's functions contain both the Hermite's polynomials and the Gaussian part), and $l_B = 1/B$ is the magnetic length. In this letter, we adopt $B=1$ unit. Following a similar procedure as above, we arrive at the following relations:

\begin{eqnarray}
&&c_{2k,n} = \frac{c^A_{k,n}+c^B_{k+1/2,n}}{\sqrt{2}}\\
&&c_{2k+1,n} = \frac{c^A_{k+1/2,n}-c^B_{k,n}}{\sqrt{2}}
\end{eqnarray}
where again $c^A_{k,n}$ and $c^B_{k+1/2,n}$ carry valid momenta, while $c^A_{k+1/2,n}$ and $c^B_{k,n}$ must be expanded in terms of other momentum ($k'$) and Landau level index ($n'$) states. Doing so, the momentum conservation is gone, and the Landau level mixing has to be considered, otherwise unitarity is violated.

Now let us consider the Hamiltonian of the following ideal cases:

{\bf 1.} $H_A$ is the Hamiltonian of the $1/3$ fermionic Laughlin state on a cylinder of length $L_y/2$, and $N_x/2$ orbital states in its lowest Landau level with Haldane's $V_1$ pseudo-potential for interaction. The length of the system along $x$ direction is given by $L_x = 2\pi N_x/L_y$ relation. We impose periodic boundary condition along $y$ direction. The Hamiltonian can be written as follows:

\begin{eqnarray}
&& H_A = \sum_{n_A = 1}^{ N_x/2}\sum_{k,l} V_{k,l}\para{L_y/2}c^{A~\dag}_{n_A}c^A_{n_A+l}c^A_{n_A+k}c^{A~\dag}_{n_A+k+l} + h.c.\\
&& p_{n_A} = 2\pi \frac{n_A}{L_y/2} = 2\pi \frac{2n_A}{L_y}.
\end{eqnarray}

where
\begin{eqnarray}
&&V_{k,l}\para{L_y/2} = A_0 \para{k^2-l^2}exp\para{-2\pi^2\frac{k^2+l^2}{\para{L_y/2}^2}} = A'_0 \para{\para{2k}^2-\para{2l}^2}exp\para{-2\pi^2\frac{\para{2k}^2+\para{2l}^2}{L_y^2}} = V_{2k,2l}\para{L_y}.\cr
&&
\end{eqnarray}

We have chosen $A_0$ and $A_1$ normalization factors such that the maximum of $V_{k,l}$ becomes 1.

{\bf 2.} $H_B$ is the Hamiltonian of the $1/3$ fermionic Laughlin state on a cylinder of length $L_y/2$, and $N_x/2$ orbital states in its lowest Landau level with Haldane's $V_1$ pseudo-potential for interaction. Again, the length of the system along $x$ direction is given by $L_x = 2\pi N_x/L_y$ relation. We impose anti-periodic boundary condition along $y$ direction. The Hamiltonian can be written as follows:

\begin{eqnarray}
&& H_B = \sum_{n_B = 1}^{N_x/2}\sum_{k,l} V_{k,l}\para{L_y/2}c^{B~\dag}_{n_B}c^B_{n_B+l}c^B_{n_B+k}c^{B~\dag}_{n_B+k+l} + h.c.\\
&&p_{n_B} = 2\pi \frac{n_B + 1/2}{L_y} = 2\pi \frac{2n_B + 1}{L_y}.
\end{eqnarray}

{\bf 3.} $H_{AB}$ is the Hamiltonian of the $1/3$ fermionic Laughlin state on a cylinder of length $L_y$, and $N_x$ orbital states in its lowest Landau level with Haldane's $V_1$ pseudo-potential for interaction. The length of the system along $x$ direction is  still given by $L_x = 2\pi N_x/L_y$ relation. We impose periodic boundary condition along $y$ direction. The Hamiltonian can be written as follows:

\begin{eqnarray}
&& H_{AB} = \sum_{n = 1}^{N_x}\sum_{k,l} V_{k,l}\para{L_y}c^\dag_{n}c_{n+l}c_{n+k}c^\dag_{n+k+l} + h.c.\\
&& p_{n} = 2\pi \frac{n}{L_y}. 
\end{eqnarray}

The Hamiltonian of an infinite genon with sharp boundaries along $y$ direction and $\beta_y$ strength is $\beta_y H_{AB} + \alpha_y\para{H_{A} + H_{B}}$, where as before $\alpha_y + \beta_y = 1$. However, we must note that the momentum states of each Hamiltonian is defined in a different basis, and to have a sharp edge genon we have to choose one basis and write them all in that basis. On the other hand, as we noted before, doing so will require both Landau level mixing and losing momentum conservation which makes computations very challenging if not quite impossible. 

To resolve the above issue, we take a different route and use the fact that the position of the single particle states along $x$ direction is tied to their momentum along $y$ direction for a given Landau level. For example, a single particle wave function  with momentum $p_n = 2\pi n/L_y$ on the lowest Landu level has a peak at $x = p_nl_B^2$. We consider a genon extended between orbitals $n_1$ and $n_2$  (recall that $\braket{x_i} = 2\pi n_i/L_y$). We also assume $\alpha_y$ and $\beta_y$ have momentum dependence which can be translated into smooth $x$ dependence. With these assumption, we simply ignore that the momentum states are note quite equal in different states and assume $c_{2k,0} = c^A_{k,0}$, and $c_{2k+1,0} = c^B_{k+1/2,0}$. This for sure means the model Hamiltonian is slightly different, but through various comparisons with our results for genons on lattice that have sharp boundaries as well as the fulfillment of strong theoretical expectations and constraints on the results, and more importantly acceptable agreements with vertical genons that can be modeled unambiguously (see the next section), we claim the two models are smoothly connected and belong to the same topological universality class. 

To summarize, our model Hamiltonian for {\em horizontal} genons with $\beta_y(n)$ strength around the $n$-th orbital is given by the following Hamiltonian:

\begin{eqnarray}
H_{\rm tot} = \sum_{n = 1}^{N_x} \overline{V}_{k,l}\para{L_y} c_{n}^\dag c_{n+l} c_{n+k}c_{n+k+l}^\dag + h.c.
\end{eqnarray}

\begin{eqnarray}
\overline{V}_{k,l}\para{L_y}  =\left\{ \begin{array}{cc}
V_{k,l}\para{L_y} & \mbox{both $k$ and $l$ are even} \\
\beta_y\para{n}V_{k,l}\para{L_y} & \mbox{either $k$ or $l$ is odd} 
\end{array}\right.
\end{eqnarray}

We would like to emphasize that above issues are absent for vertical infinite genons, and the fact that our energy spectrum results for horizontal genons agree quite well with those for vertical genons is another signature that our model Hamiltonian of genons for FQH state captures all the essential physics needed for genons.

\begin{figure}[!t]
\centering
\centerline{\includegraphics[width=0.6\linewidth]{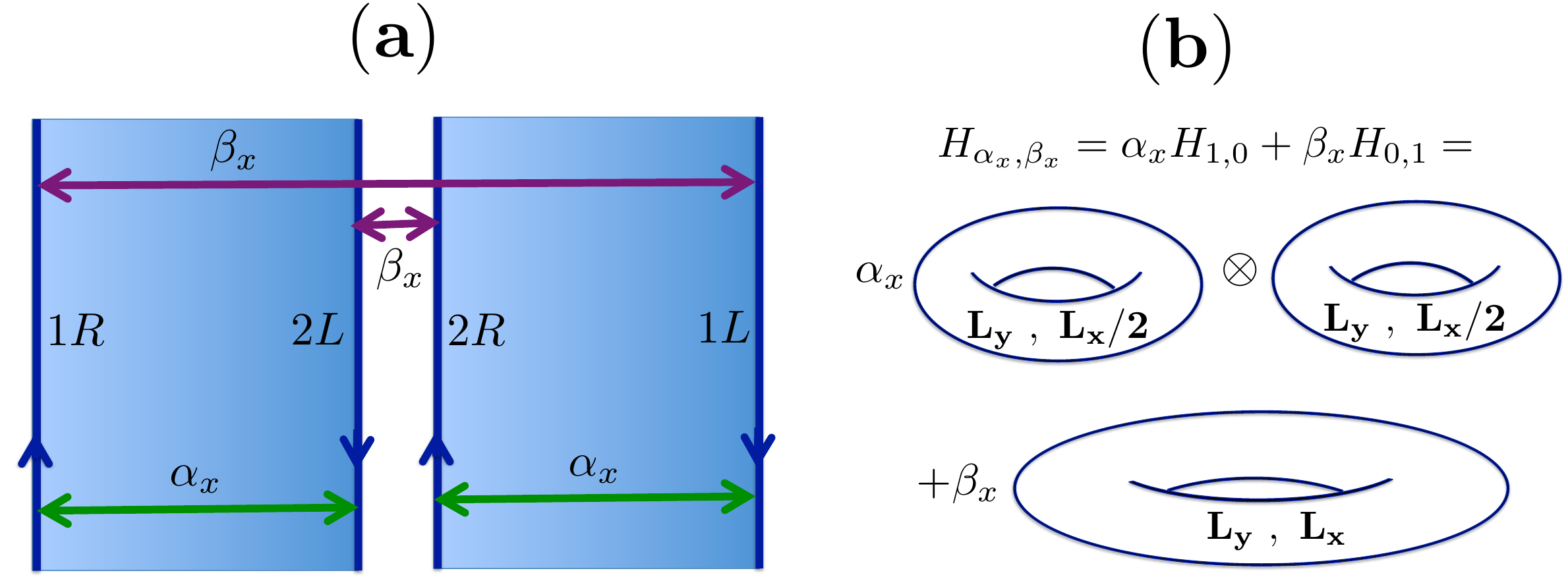}}
\caption{A schematics of a vertical infinite genon (a) and the corresponding Hamiltonian and space manifold (5).}
\label{Appendix_Fig_02}
\end{figure}

\section*{B. Vertical infinite genons and GSD calculations}
In this section we comment on our model of genons extended all the way to the boundaries along $y$ axis  for FQH systems. We impose PBC along $y$ direction. Similar to the case of horizontal genons, we need to create four edge states first, two of which are chiral and the remaining two ones are anti-chiral. To this end, we must carve four edges as shown in Fig. \ref{Appendix_Fig_02}(a). Then, in general the edge 1R can couple to both 1L and 2L, with coupling strengths equal to $\beta_x$ and $\alpha_x=1-\beta_x$, respectively. We assume the same couplings couple 2L to 2R and 1R, respectively. Now, it is easy to verify that the Hamiltonian of this genon is a linear combination of the Hamiltonian of the two extreme cases, as shown in Fig. \ref{Appendix_Fig_02}(b), i.e., the Hamiltonians associated with $\alpha_x = 1$ ($\beta_x = 0$) and $\beta_x = 1$ ($\alpha_x = 0$). From the cut and glue approach we can easily tell that the Hamiltonian associated with $\beta_x = 1$ corresponds to gluing edge 1R to 1L and 2R to 2L, after which we obtain a single torus of length $L_x$ along $x$ and $L_y$ along $y$ direction. Moreover,  $\alpha_x = 1$ corresponds to gluing edge 1R to 2L and 2R to 1L, after which we are left with two decoupled tori each of length $L_x/2$ along $x$ and $L_y$ along $y$ direction. The Hamiltonian of the first and the second case can be easily constructed using the Haldane pseudo-potentials utilized in the previous section. Similar to the horizontal genons case, our theoretical prediction is 9-fold degeneracy for $\alpha_x > 1/2$ and 3-fold degeneracy for $\alpha_x < 1/2$ and a quantum critical point at $\alpha_x = 0$ in the thermodynamic limit. Of course, there would be finite size effect. However, as Fig. \ref{Appendix_Fig_03} indicates, the finite size effect becomes weaker by considering aspect ratios ($L_y/L_x $) closer to unity.

\begin{figure}[!t]
\centering
\centerline{\includegraphics[width=0.6\linewidth]{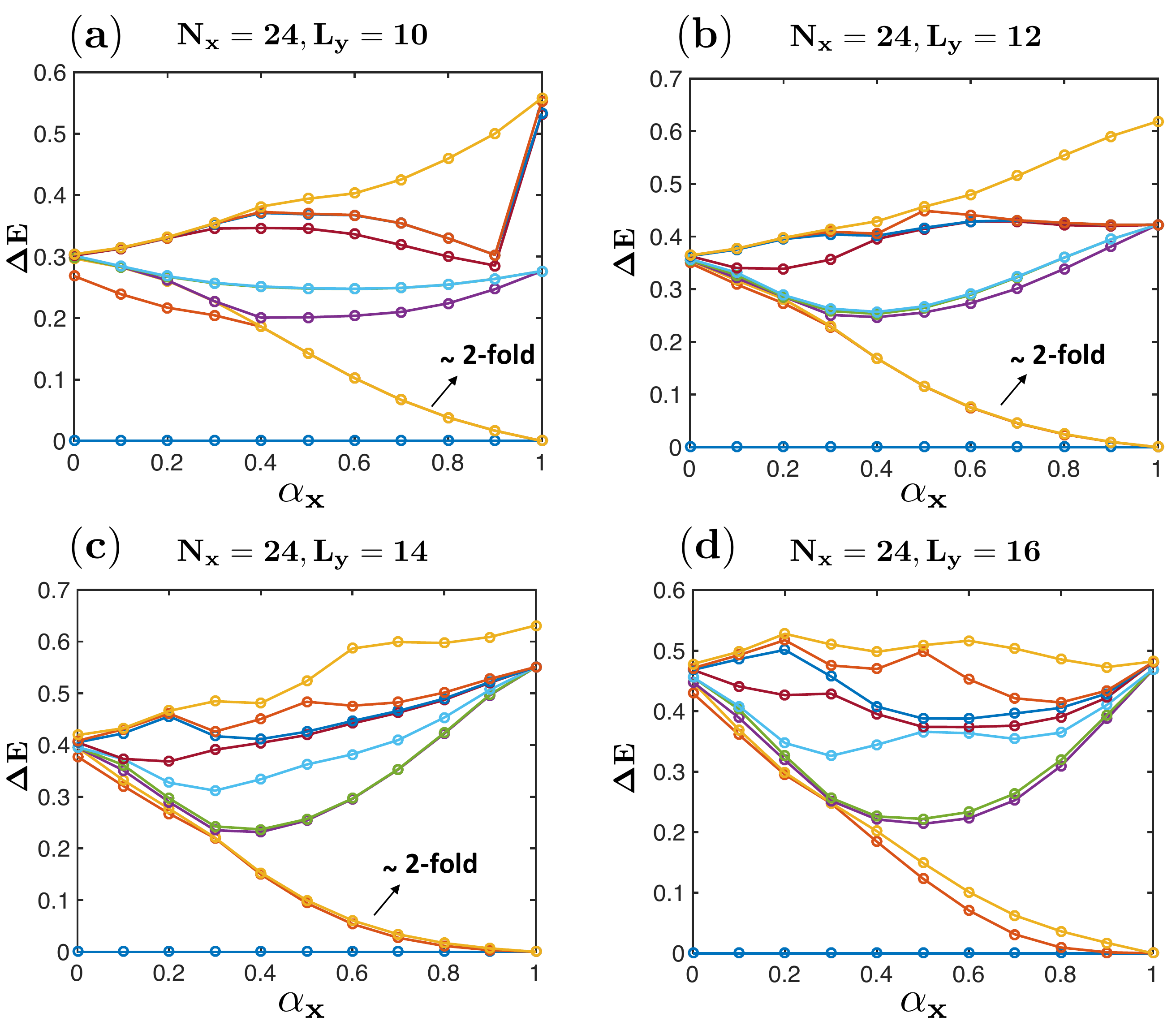}}
\caption{Lowest ten energy eigenvalues of an infinite vertical genon versus coupling strength ($\alpha_x$) as well as $L_y$. The shown energy spectrum is computed in one of the three topological sectors. The other two sector have identical energy spectra, so to find the overall degeneracy, the GSD shown in these plots must be tripled. These plots corroborates our expectation of having 9-fold GSD for $\alpha_x > 1/2$ in the thermodynamic limit.}
\label{Appendix_Fig_03}
\end{figure}

\section*{C. More results for GSD}

\begin{figure}[!t]
\centering
\centerline{\includegraphics[width=0.3\linewidth]{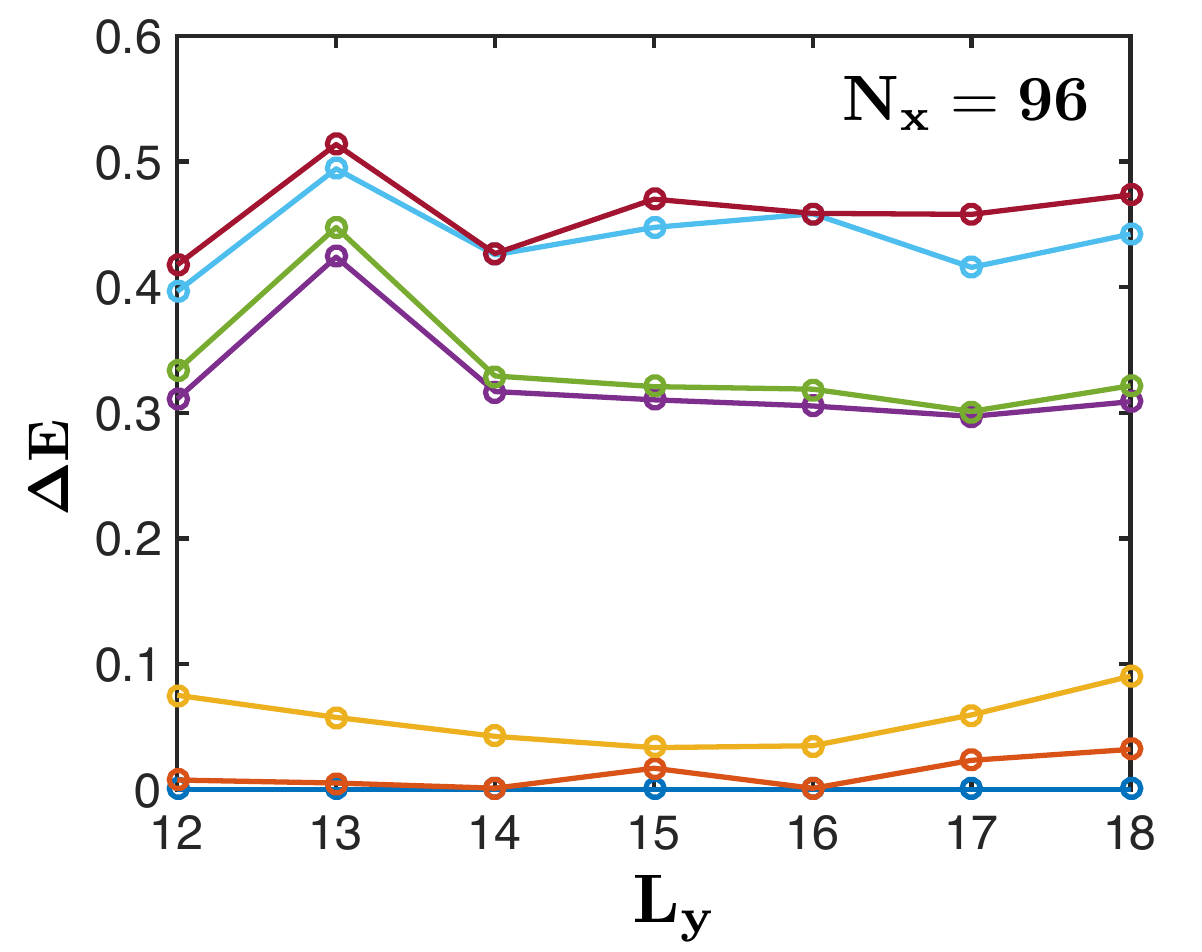}}
\caption{The energy spectrum of a finite horizontal genon (with $\alpha_y$ changing from 0 to 1) for a Landau level with 96 orbitals (single particle basis states). Again, these energies are projections to one of the three topological sectors only.}
\label{Appendix_Fig_04}
\end{figure}

\begin{figure}[!t]
\centering
\centerline{\includegraphics[width=0.3\linewidth]{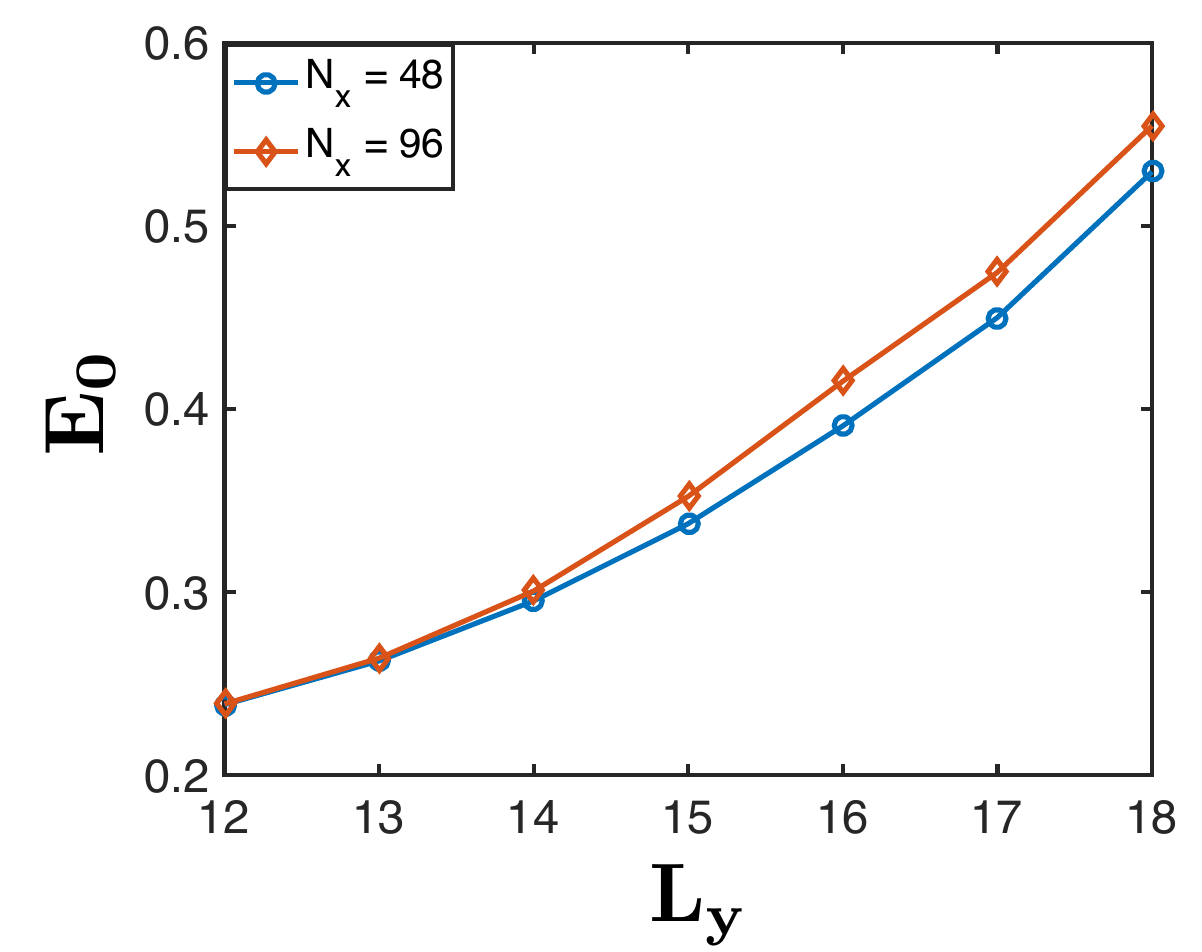}}
\caption{The ground-state energy of the finite horizontal genons (with $\alpha_y$ changing from 0 to 1) as a function of $L_y$ for two system sizes: $N_x=48$ and $N_x = 96$ orbitals. The ground-state energy is almost independent of the system size, which points toward the localization of PFZMs bound to the domain walls. }
\label{Appendix_Fig_05}
\end{figure}

In this section we provide more results for the energy spectrum of both horizontal and vertical genons (see Fig. \ref{Appendix_Fig_04}). We show that for $N_x = 96$ we obtain energy spectrum of a finite horizontal genon extended between $x_1 = - L_x/4$ and $x_2 = L_x/4$ which are similar to Fig. 2 of the main text. Additionally, we observe that the ground-state energy is almost independent of the genon's size, which implies the localization of the PFZMs (see Fig. \ref{Appendix_Fig_05}).

\section*{D. The central charge measurement}

In this section we provide the entanglement entropy of two infinite horizontal genons with $\alpha_y = 0$, and $\alpha_y=1/2$ parameters, respectively. Using them, we can extract the central charge of the critical region associated with $\alpha_y=1/2$. We obtain similar plots for other parameters used in Fig. 4 of the main text.

\begin{figure}[!t]
\centering
\centerline{\includegraphics[width=0.6\linewidth]{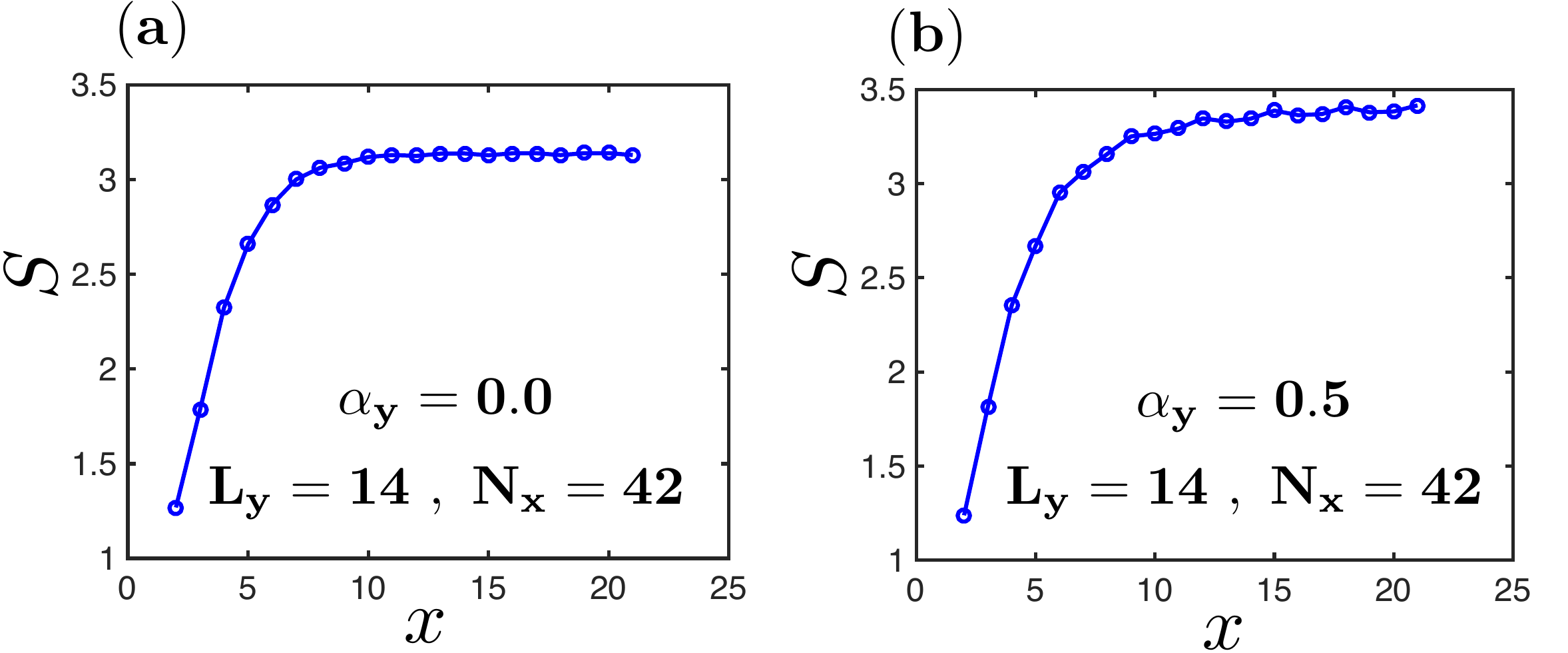}}
\caption{The von Neumanns entanglement entropy calculated for $\alpha_y = 0$ (a) and $\alpha_y = 1/2$ genons on the torus geometry. Their difference was used in the main text to extract the central charge of critical genons.}
\label{Appendix_Fig_05}
\end{figure}

\end{widetext}

\end{document}